\documentclass[10pt,twocolumn,letterpaper]{article}

\usepackage{cvpr}

\usepackage{graphicx}
\usepackage{booktabs}
\usepackage{amsmath,amssymb}
\usepackage{multirow}
\usepackage{float}
\graphicspath{{figures/}}

\definecolor{cvprblue}{rgb}{0.21,0.49,0.74}
\usepackage[pagebackref,breaklinks,colorlinks,allcolors=cvprblue]{hyperref}
\AtBeginDocument{%
  \crefname{figure}{Fig.}{Figs.}%
  \Crefname{figure}{Figure}{Figures}%
}

\def\paperID{*****}
\def\confName{AAAI}
\def\confYear{2027}

\title{Shape-Based Inductive Bias for Glioma Grading from Tumor Contours}

\author{%
  Puneet Velidi\\
  Department of Mathematics and Statistics\\
  University of Victoria\\
  {\small \texttt{puneet.velidi@gmail.com}}
  \and
  Michelle F. Miranda\\
  Department of Mathematics and Statistics\\
  University of Victoria\\
  {\small \texttt{michellemiranda@uvic.ca}}
  \and
  Farouk Nathoo\\
  Department of Mathematics and Statistics\\
  University of Victoria\\
  {\small \texttt{nathoo@uvic.ca}}
  \and
  Ashery Mbilinyi\\
  Department of Computer Science\\
  University of Victoria\\
  {\small \texttt{ashery@uvic.ca}}
  \and
  C\'edric Beaulac\\
  D\'epartement de math\'ematiques\\
  Universit\'e du Qu\'ebec \`a Montr\'eal\\
  {\small \texttt{beaulac.cedric@uqam.ca}}
}

\begin{document}
\maketitle

\begin{abstract}
Glioma grading from tumor contours is often treated as a pixel problem even when the signal of interest is shape. We align closed contours with a functional shape-alignment framework, separate global deformation from residual Fourier shape, and organize these quantities as frequency-ordered tokens. In five-fold patient-disjoint cross-validation on BraTS~2020 tumor contours, a compact multilayer perceptron (MLP) achieves the highest mean balanced accuracy at 71.5\%, compared with 65.9\% for ResNet-18 and 63.3\% for ViT-Tiny. It also gives the highest mean low-grade glioma F1 at 54.9\%. Its pooled out-of-fold balanced accuracy is 72.4\% (patient-bootstrap 95\% CI: 66.4--77.8\%). The selected MLPs use 2.9k--117.3k parameters across folds, at least 46 times fewer than the pixel baselines. In a controlled noise-free simulation testing labels derived from harmonic and deformation features, shape-based models reach 56.3--71.5\% balanced accuracy while the pixel models remain near chance at 50.0--52.5\%. This work demonstrates how incorporating a shape-based inductive bias at the representation level can enable substantial dimensionality reduction and increased performance while also improving interpretability.
\end{abstract}

\section{Introduction}
\label{sec:intro}

Object shape is often the signal of interest in scientific imaging. Tumor margins, cell outlines, and organ silhouettes are all examples of where labels depend on geometry. However, the default deep-learning formulation typically rasterizes the object such that pixel-based image models have to indirectly learn morphological features while also learning invariance to translation, rotation, and scale; in small labelled cohorts, this spends statistical power on variation that a user may already have prior information about.

Classical shape analysis has traditionally addressed this issue by first aligning objects and then classifying or comparing them in a coordinate system where nuisance transformations have been removed, using tools such as Fourier descriptors, Procrustes alignment, and square-root velocity function (SRVF) representations~\cite{zahn1972fourier,kuhl1982elliptic,dryden1998statistical,srivastava2011shape,srivastava2016functional}. The machine-learning analogue is a shape-based inductive bias, in which curve geometry supplies a quotient space, the space of shapes with translation, rotation, scale, and starting point factored out, before supervised learning begins, in the same spirit as geometric and equivariant learning methods that build symmetry structure into the representation~\cite{cohen2016group,bronstein2021geometric}.

We study this representation question in glioma grading from BraTS~2020 tumor masks~\cite{brats2020,menze2015brats}, an application in which radiologists and radiomics pipelines have long treated tumor morphology as clinically meaningful~\cite{lambin2012radiomics,aerts2014decoding,gillies2016radiomics}. High-grade gliomas (HGGs) often present with larger and more irregular margins, while low-grade gliomas (LGGs) tend to be smaller and more compact~\cite{schaefer2013morphology}.

\begin{figure}[t]
    \centering
    \includegraphics[width=\linewidth]{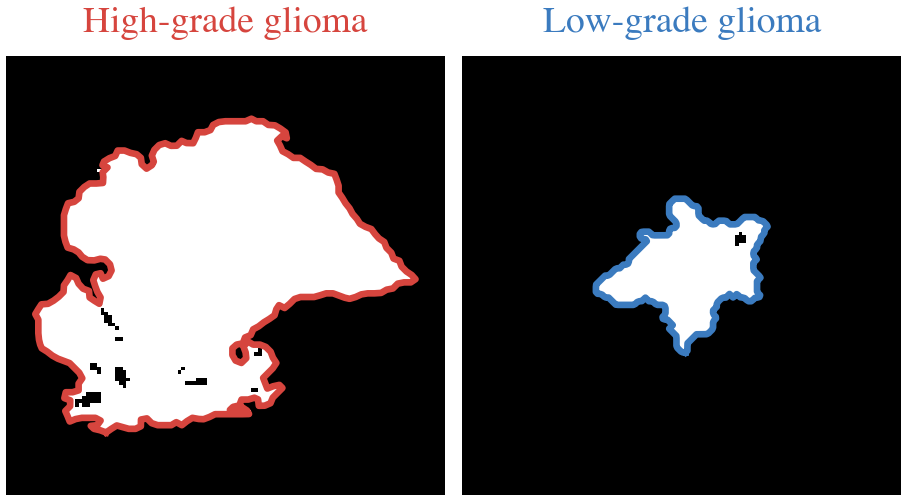}
    \caption{Tumor contours extracted from BraTS 2020 segmentation masks. High-grade gliomas (left) tend to be larger with more irregular boundaries; low-grade tumors (right) are typically smaller and more compact. These examples motivate a contour-level representation for morphology classification.}
    \label{fig:contours}
\end{figure}

The technical basis for our paper is the functional-shape alignment framework introduced by Moindji\'e \etal~\cite{moindjie2025functional}. Here, a closed contour is treated as a deformation of a latent shape under translation, scale, rotation, and cyclic reparameterization. In a Fourier basis, the starting-point shift becomes a sparse orthogonal block transform, making joint phase-and-rotation alignment a finite-dimensional Procrustes problem. We retain the estimated deformation variables as model inputs alongside the residual Fourier coefficients described in Section~\ref{sec:tokens}.

We use the term shape token for either the deformation vector or one harmonic residual vector. Lower harmonics encode coarse geometry such as elongation and asymmetry, while higher harmonics encode lobes, indentations, and local boundary roughness. This representation allows us to compare a Shape Transformer, which updates each token using the other deformation and harmonic tokens through self-attention, with a compact MLP that operates on the flattened token representation. In five-fold patient-disjoint cross-validation, the MLP attains the highest mean balanced accuracy and LGG F1. Its selected configurations use 2.9k--117.3k parameters, compared with 0.80M--4.75M for the Shape Transformer and 5.43M--11.2M for the pixel baselines.

This minority-class result suggests greater statistical efficiency. In the imbalanced HGG/LGG setting, overall accuracy can be dominated by HGG cases even when LGG tumors are poorly identified. The compact Fourier representation allows a small MLP to remain competitive and even superior on LGG detection without a high-capacity classifier.

\paragraph{Contributions.}
\begin{enumerate}
    \item We formulate aligned Fourier contours as \emph{shape tokens}, with one six-value deformation token and one six-value residual shape token per harmonic. With $K=32$, the representation has $(K+1)\times6=198$ scalar inputs.
    \item We show on BraTS~2020 that compact MLPs over at most 198 scalar inputs attain the highest mean balanced accuracy and LGG F1 under patient-disjoint cross-validation. The selected MLPs use 2.9k--117.3k parameters, at least 46 times fewer than the pixel baselines with at least 2,000 times less training time.
    \item We use a controlled simulation to examine how global deformation and mid-frequency shape information affect shape-based and pixel-based classifiers applied to the same generated objects.
\end{enumerate}

\section{Related Work}
\label{sec:related}

\paragraph{Radiomics and medical shape.}
Radiomics encodes clinical images through quantitative features and has repeatedly found morphology to be informative for tumor phenotype and outcome~\cite{lambin2012radiomics,aerts2014decoding,gillies2016radiomics}. Prior glioma-grading studies have combined shape with intensity or texture features~\cite{zacharaki2009classification,cho2018glioma}, while CNN pipelines have used the largest axial slice or full 3D MRI~\cite{zhuge2020automated,xu2022automated}. Modern segmentation networks such as U-Net~\cite{ronneberger2015unet}, nnU-Net~\cite{isensee2021nnunet}, and UNETR~\cite{hatamizadeh2022unetr} produce high-quality masks. Our analysis begins once a mask has been obtained and uses the largest-area axial slice, as described in Section~\ref{sec:tasks}. This paper studies the value of transforming that mask into a shape coordinate system before classification.

\paragraph{Classical shape representations.}
Classical contour representations include Fourier descriptors~\cite{zahn1972fourier,kuhl1982elliptic,persoon1977shape}, chain codes~\cite{freeman1961encoding}, shape contexts~\cite{belongie2000shape,belongie2002shape}, active shape models~\cite{cootes1995active}, and Procrustes or elastic shape analysis~\cite{dryden1998statistical,rohlf1990morphometrics,bookstein1997morphometric,srivastava2011shape,srivastava2016functional}. Contemporary shape analysis often uses the SRVF framework, which represents curves through velocity-normalized functions and compares shapes after quotienting rotation and reparameterization~\cite{srivastava2011shape,srivastava2016functional}. These methods make invariance explicit by aligning shapes before comparing them. In statistical shape analysis, downstream classifiers are often intentionally simple. Aligned shapes are projected to tangent, principal-component, or basis coordinates and then passed to functional linear or component-based classifiers. Moindji\'e, Beaulac, and Descary develop a functional view of curve alignment and latent-shape estimation~\cite{moindjie2025functional}, with an application to functional classification of multivariate planar X-ray curves~\cite{moindjie2025multivariate}. Inspired by that alignment framework, we ask whether a compact nonlinear classifier can learn from the resulting coefficients.

\paragraph{Functional neural representations.}
A small but growing body of work has integrated neural networks into functional data analysis (FDA), including deep-learning models with functional inputs~\cite{thind2023deep_functional_inputs}, functional outputs~\cite{wu2023neural_functional_output}, and autoencoder-based approaches to smoothing and representation learning~\cite{wu2024functional_autoencoder}. This literature motivates a middle ground between classical functional linear models and large pixel-based networks, where nonlinear decision rules can be learned without giving up a functional coordinate system whose axes retain geometric meaning.

\paragraph{Tokenized learning.}
Transformers~\cite{vaswani2017attention} and BERT-style classification~\cite{devlin2019bert} motivate learned pooling over token sequences. Vision Transformers~\cite{dosovitskiy2021image} tokenize images into patches, leaving geometric invariance largely to the data and model capacity. Transformer-based models have also been used when the input is naturally a collection of elements, including set-structured data and point clouds~\cite{lee2019set,zhao2021point}. In our case, the tokens are frequency-ordered coordinates of an aligned contour. Because harmonic bands correspond to different, ordered scales of morphology, self-attention could allow the classifier learn interactions among deformation variables, low-frequency elongation and asymmetry, and higher-frequency boundary detail.

\subsection{Shape Inductive Bias}
\label{sec:bias}

We introduce a shape-based inductive bias at the representation level through a functional basis representation of tumour contours. This contrasts with CNNs, which are often biased toward classification using texture rather than shape \cite{Geirhos2018ImageNettrainedCA,Zhang2023ConvolutionalNN} and are not necessarily translation invariant \cite{Biscione2021ConvolutionalNN}. Increasing the amount of shape information available to a model has also been shown to improve generalizability and robustness to perturbations and some forms of distribution shift. Geirhos \etal~\cite{Geirhos2018ImageNettrainedCA}, for example, augment the training data to induce a stronger shape bias by suppressing texture while preserving object shape. Our approach can be viewed as a more direct version of this idea, since we entirely supress texture.

\section{Tasks}
\label{sec:tasks}

\subsection{BraTS 2020}

 We use BraTS~2020 expert segmentation masks~\cite{brats2020,menze2015brats} to construct a
  morphology-derived binary target from the tumor subregions present in each extracted component.
  Components containing the complete triad of necrotic or non-enhancing tumor core, peritumoral
  edema, and enhancing tumor (labels 1, 2, and 4) are assigned HGG. Their co-occurrence reflects
  three complementary features of aggressive disease which are tissue necrosis, disruption of the blood--brain
  barrier with contrast enhancement, and an extensive infiltrative or edematous response. The remaining
  observed combinations are assigned LGG because they lack this 
  heterogeneous signature and represent comparatively lower-grade-like morphology. Thus,
  the target is an operational imaging-phenotype classification rather than a histopathologic
  diagnosis~\cite{schaefer2013morphology}, which suffices for our demonstration of shape-based classification. The export contains 413 contours from 368 patients,
  comprising 318 HGG and 95 LGG contours. Of these patients, 331 contribute one contour and 37
  contribute multiple contours. We focus on the minority LGG class because accurate detection must exceed majority-class
  recognition and may identify disease before progression to less treatable HGG.

\paragraph{From 3D volumes to 2D contours.}
For each patient we select the axial cross-section with the largest total tumor area and binarize the whole-tumor mask. Each connected component with at least 50 pixels is treated as a separate tumor and converted to a level-set contour with the marching-squares implementation in scikit-image~\cite{lorensen1987marching,skimage2026findcontours}. The largest closed boundary within each component is kept. Marching squares produces contours with different numbers and spacings of vertices, so we resample each boundary to $P{=}256$ points uniformly along arc length before Fourier projection. The coefficient-space formulation no longer depends on the original vertex count after this common projection~\cite{moindjie2025functional}. Finite resampling and truncation can still introduce aliasing, which is one reason we use uniform arc-length sampling before retaining $K\leq32$ harmonics. In the simulation, noise is added to the 256 generated vertices and both tokenization and rasterization use those same noisy vertices without another resampling step.

\paragraph{Splits and metric.}
The reported BraTS numbers use five-fold outer cross-validation grouped by patient and stratified by the component-level binary label. The outer folds function as test sets: each contour appears in exactly one outer test fold, and no patient appears in both outer training and test data within a fold. Within each outer-training fold, the first split from a four-fold patient-grouped partition provides inner training and validation subsets. We select token bandwidth, architecture, hyperparameters, and training epoch from validation balanced accuracy; the selected setup is then retrained from scratch on the complete outer-training fold and evaluated once on the outer test fold. All models use the same outer and inner partitions. We report fold means and sample standard deviations and additionally pool all 413 out-of-fold predictions. Confidence intervals are obtained by resampling patients together with all of their contours.

\subsection{Controlled Simulation}
\label{sec:simulation_setup}

We use our simulation to investigate how shape-based classifiers use deformation and latent-shape information in a controlled geometric setting. We draw a latent shape indicator $\mathrm{cls}\sim\mathrm{Bernoulli}(0.5)$ independently of the deformation variables. This indicator controls the added harmonics but is not the target label $y$. For $t\in[0,2\pi)$, the planar base curve is $\mathbf{C}_0(t)=r(t)(\cos t,\sin t)$, where
\begin{equation}
    r(t) = 1 + c_3 \cos(3t+\phi_3)
           + \mathbf{1}\{\mathrm{cls}=1\}\sum_{k=4}^{6} b_k \cos(kt+\phi_k),
\end{equation}
Here $c_3 \sim U(0.18,0.32)$ and $\phi_3 \sim U(0,2\pi)$ randomize the shared three-lobed base shape. For class 1 only, the amplitudes $b_k \sim U(0.020,0.030)$ add small bumps at harmonics $k=4,5,6$, with independent random phases $\phi_k$. The base curve is then transformed by isotropic scale $s=\exp(u)$, $u\sim U(\log 0.3,\log 3.0)$, small shear, arbitrary rotation, and translation $(t_x,t_y)\sim U(-1.5,1.5)^2$.

Labels deliberately combine information from the deformation and latent-shape components,
\begin{equation}
    \begin{aligned}
    \eta &= \log s + 0.5\,\mathrm{sign}(t_x)
            + 1.2(\mathrm{cls}-0.5) + \epsilon,\\
    y &= \mathbf{1}\{\eta > 0\},
    \qquad \epsilon \sim \mathcal{N}(0,0.1).
    \end{aligned}
\end{equation}
The terms $\log s$ and $\mathrm{sign}(t_x)$ are available to a shape-based classifier through the deformation token, while $\mathrm{cls}$ is encoded by the mid-frequency residual shape tokens. Pixel-based classifiers read a $64{\times}64$ filled raster produced after recentering and rescaling the polygon, which are common preprocessing steps. To simulate noise, independent Gaussian noise was added to each coordinate of every contour vertex, ($\widetilde{\mathbf c}_i=\mathbf c_i+
  \boldsymbol{\epsilon}_i$), where ($\boldsymbol{\epsilon}_i\sim\mathcal N(\mathbf 0,\sigma_c^2\mathbf I_2)$) and ($\sigma_c\in{0,0.10}$). The same perturbed contour was used for Fourier tokenization and rasterization, while the label remained unchanged and no pixel-level noise was added.

\paragraph{Sample sizes, noise, and bandwidth.}
The reported simulation aggregate uses tokenizer bandwidth $K{=}32$, selected by the procedure in Section~\ref{sec:sweep}, sample sizes $N \in \{50,100,250,1{,}000\}$, and contour-noise levels $\sigma \in \{0,0.10\}$. Here ``noise-free'' means $\sigma=0$. At $\sigma=0.10$, independent Gaussian noise is added to contour vertices before either tokenization or rasterization. Architectures are selected once on a separate noise-free validation set and then held fixed for both test conditions. Final estimates use 20 independent training runs and the same fixed 2,000-sample test set within each noise condition which prevents model selection from adapting to a particular test-noise level. This also means the noisy condition measures robustness away from the selection/validation distribution.

\section{Shape-Based Method}
\label{sec:methods}

The same processing and tokenizaiton pipeline is shared by the BraTS analysis and the simulation. For efficient batched computation, a segmentation boundary or simulated contour is first represented on a uniform 256-point grid and projected once to a common Fourier basis. Alignment and token construction then operate in coefficient space, and either a compact MLP or a Shape Transformer performs classification.

\begin{figure}[t]
    \centering
    \includegraphics[width=\columnwidth]{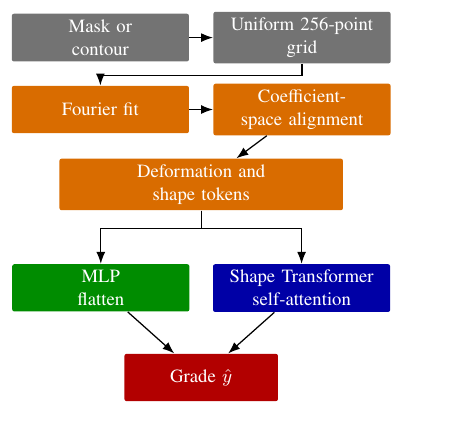}
    \caption{Contour-token pipeline. The contour is converted to a deformation token and frequency-ordered residual shape tokens. Downstream shape-based classifiers receive the same token matrix, the MLP flattens it, while the Shape Transformer mixes harmonic information through self-attention.}
    \label{fig:pipeline}
\end{figure}

\subsection{Functional-Shape Alignment and Tokenization}
\label{sec:shapefda}

Inspired by Moindji\'e \etal's functional formulation~\cite{moindjie2025functional}, each observed planar contour $\mathbf{C}(t)=(X(t),Y(t))$, $t\in[0,1]$, is treated as a deformation of a latent shape,
\begin{equation}
\label{eq:shape_model}
\mathbf{C}(t) = \rho\, \mathbf{O}_\theta\, \tilde{\mathbf{C}} \circ \gamma_\delta(t) + \mathbf{T},
\end{equation}
with scale $\rho>0$, rotation $\mathbf{O}_\theta$, translation $\mathbf{T}$, and cyclic starting-point shift $\gamma_\delta(t)=\mathrm{mod}(t-\delta,1)$. We use a training-derived reference template $\boldsymbol{\mu}$. During inner model selection it is the first preshape in the deterministic inner-training order; after selection it is recomputed as the first preshape in the complete outer-training fold. Validation and outer-test contours are aligned to this training reference and never contribute to its estimation. This fixed-reference choice differs from iterative Fr\'echet-mean estimation in the full functional-shape procedure~\cite{moindjie2025functional}, and sensitivity to the reference choice remains to be tested.

\paragraph{Preshape.}
Our implementation estimates translation and scale from the uniformly resampled points,
\begin{equation}
\begin{aligned}
    \mathbf{T}_i &= P^{-1}\sum_{j=1}^{P}\mathbf{C}_{ij},\\
    \rho_i &= \left(P^{-1}\sum_{j=1}^{P}
    \|\mathbf{C}_{ij}-\mathbf{T}_i\|_2^2\right)^{1/2}.
\end{aligned}
\end{equation}
The preshape is $\mathbf{C}_i^*=(\mathbf{C}_i-\mathbf{T}_i)/\rho_i$. This point-space calculation was used to share a simple vectorized preprocessing interface across BraTS contours and simulated curves. It is not required by Moindji\'e \etal's functional approach. We then compute the Fourier coefficients of the uniformly sampled preshape using
\begin{align}
    a_k &= \frac{2}{P}\sum_{j=0}^{P-1} x_j\cos(2\pi kj/P), \notag\\
    b_k &= \frac{2}{P}\sum_{j=0}^{P-1} x_j\sin(2\pi kj/P),
\end{align}
and analogously for the $y$ coordinate.

\paragraph{Rotation and phase.}
For each candidate $\delta$ on a 101-point grid, the Fourier shift property applies the corresponding block-diagonal rotation to each harmonic. The grid gives a starting-point resolution of about one percent of the contour and was chosen as a practical accuracy and cost tradeoff. It can introduce phase quantization below one grid step. The rotation $\theta$ is then solved in closed form by the Procrustes trace objective
\begin{equation}
    (\hat\theta_i,\hat\delta_i)=
    \arg\min_{\theta,\delta}
    \left\|\boldsymbol{\alpha}_i\,\boldsymbol{\beta}(\delta)-\mathbf{O}_\theta \boldsymbol{\mu}\right\|_F^2,
    \label{eq:fda_objective}
\end{equation}
where $\boldsymbol{\alpha}_i$ stacks the preshape Fourier coefficients and $\boldsymbol{\beta}(\delta)$ is the sparse orthogonal phase-shift matrix~\cite{moindjie2025functional}. The implementation evaluates the grid, keeps the best objective, and returns $(\hat\theta_i,\hat\delta_i)$.

\paragraph{Coefficient-space implementation.}
Following Moindji\'e \etal, phase, inverse rotation, and scale are applied directly to the fitted coefficients through the block structure of the basis~\cite{moindjie2025functional}. Supplementary Section~C compares this implementation with an equivalent spatial-roundtrip approximation and shows identical aggregate metrics in a fixed-split BraTS implementation check.

\subsection{Token Layout}
\label{sec:tokens}

The default token matrix is $\mathbf{X}_i\in\mathbb{R}^{(K+1)\times 6}$.

\paragraph{Deformation token.}
The first token stores the functional-shape deformation variables
\begin{equation}
    (T_x,T_y,1/\rho,\delta,\theta,0),
\end{equation}
where the first five values are standardized using means and standard deviations fitted only on the relevant training partition, then applied unchanged to validation or outer-test contours. During final outer-fold training, these statistics are refitted on the complete outer-training fold. The padding value remains zero. Thus, test examples neither influence one another nor contribute distributional information to preprocessing. Unlike a learned \texttt{[CLS]} token, the deformation token carries physical contour information and also serves as the position-0 readout for the shape-based models.

\paragraph{Shape tokens.}
Each shape token stores the residual coefficients and amplitudes for one harmonic,
\begin{align}
    \mathbf{v}_{ik} &=
    (a_{xk},b_{xk},a_{yk},b_{yk},A_{xk},A_{yk})_i,\\
    \mathbf{z}_{ik} &=
    \mathrm{zscore}_{\mathrm{rows}}\!\left[
    \mathbf{v}_{ik} - \mathbf{v}_{\mu k}
    \right].
\end{align}
where $A_{xk}=\sqrt{a_{xk}^2+b_{xk}^2}$ and $A_{yk}=\sqrt{a_{yk}^2+b_{yk}^2}$. These amplitudes make harmonic magnitude directly available while the signed coefficients retain phase information.
Low-order tokens encode coarse geometry; higher-order tokens encode boundary lobes and local detail. With $K{=}32$, the largest token representation has $33{\times}6=198$ scalars. Bandwidth is selected independently within each outer fold for each architecture over $K\in\{8,16,24,32\}$. The selected models can therefore receive different harmonic bandwidths, so their comparison is not a controlled capacity comparison on identical inputs.

\subsection{Classifiers}
\label{sec:classifiers}

All shape-based classifiers consume the same token matrix for a given $K$.

\paragraph{MLP.}
The token matrix is flattened and passed through two GELU hidden layers,
\begin{align*}
\mathbf{X} &\xrightarrow{\mathrm{flatten}} \mathbb{R}^{6(K+1)}
           \xrightarrow{\mathrm{Linear+GELU+drop}} \mathbb{R}^{d_h},\\
\mathbb{R}^{d_h} &\xrightarrow{\mathrm{Linear+GELU+drop}} \mathbb{R}^{d_h}
           \xrightarrow{\mathrm{Linear}} \hat{y}.
\end{align*}
There is no positional embedding and no parameter sharing across harmonics. Across the five BraTS folds, the selected configurations span $K\in\{8,16,32\}$, $d_h\in\{32,128,256\}$, and dropout in $\{0,0.1,0.2\}$, for \textbf{2.9k--117.3k} parameters.

\paragraph{Shape Transformer.}
The transformer is a pre-norm encoder~\cite{vaswani2017attention} with a linear token projection, learned positional embeddings, $L$ self-attention blocks with $H$ heads, GELU feed-forward layers of width $\max(64,4d)$, and position-0 readout. Self-attention updates are
\begin{equation}
    \alpha_{ij} = \frac{\exp(\mathbf{q}_i^\top \mathbf{k}_j / \sqrt{d_k})}{\sum_m \exp(\mathbf{q}_i^\top \mathbf{k}_m / \sqrt{d_k})},\quad
    \mathbf{h}_i' = \sum_j \alpha_{ij} \mathbf{v}_j.
\end{equation}
Across the five BraTS folds, the selected configurations span $K\in\{8,16,24,32\}$, $d\in\{128,192,256\}$, $H\in\{4,8\}$, and $L\in\{4,6\}$, for \textbf{0.80M--4.75M} parameters.

\paragraph{Pixel baselines.}
BraTS pixel-based baselines are ResNet-18~\cite{he2016deep} and ViT-Tiny~\cite{dosovitskiy2021image} on one-channel $224{\times}224$ rasterized tumor masks. ResNet-18 uses its standard $7{\times}7$, stride-2 stem, and ViT-Tiny uses $16{\times}16$ patches. Both are trained from scratch. Simulation pixel-based baselines uses the same architectures on $64{\times}64$ filled rasters.

\paragraph{Training.}
BraTS runs train with AdamW, learning rate $10^{-3}$, weight decay $10^{-2}$, batch size 16, 40 epochs, and inverse-square-root class-balanced sampling. The simulation token runner uses Adam, learning rate $2{\times}10^{-3}$, batch size 64, 60 epochs, and no early stopping. The pixel runner uses the same learning rate and batch size for 30 epochs. These fixed budgets are not compute matched, which limits architecture-level comparisons in the simulation.

\section{Model Selection Protocol}
\label{sec:sweep}

 The outer BraTS folds are the test partitions used to estimate out-of-sample performance. Within each outer-training fold, model selection uses one patient-grouped inner validation split. Simulation selection is performed on a noise-free holdout set. Final simulation tables hold selected configurations fixed across random initializations and noise levels.

\paragraph{BraTS token-model sweep.}
The BraTS sweep uses the Fourier-space token exporter for $K \in \{8,16,24,32\}$. MLP widths are $d_h \in \{32,64,128,256,512\}$ with dropout $p \in \{0.0,0.1,0.2\}$. The Shape Transformer uses $d \in \{128,192,256\}$, $H \in \{4,8\}$, and $L \in \{4,6\}$ subject to $H \mid d$. For every candidate, the reference contour and deformation-feature normalization are fitted on the inner-training subset and applied unchanged to its validation subset. Hyperparameters and training epochs are selected using one grouped inner validation split, after which each model is retrained. After selection, preprocessing is refitted on the full outer-training fold before its test contours are transformed. BraTS results are means over five patient-disjoint outer folds, which serve as the test partitions. Table~\ref{tab:brats_token_grid} summarizes the configurations selected independently within the five outer folds. Outer-fold results are reported separately in Table~\ref{tab:brats_acc}.

\begin{table}[t]
\centering
\small
\begin{tabular}{lcc}
\toprule
Model & Selected $K$ & Params \\
\midrule
MLP & 8, 8, 8, 16, 32 & 2.9k--117.3k \\
Shape Trans. & 8, 16, 24, 32, 32 & 0.80M--4.75M \\
\bottomrule
\end{tabular}
\caption{BraTS 2020 token configurations selected by grouped inner-validation balanced accuracy. One configuration is selected within each outer fold; $K$ values are sorted for compactness. MLP widths span 32--256; transformer widths span 128--256, heads span 4--8, and layers span 4--6.}
\label{tab:brats_token_grid}
\end{table}

The selected model varies materially across folds for both token models, which is expected in this small dataset. The MLP nevertheless remains compact in every fold, with its largest selected configuration still at least 46 times smaller than ViT-Tiny. Because bandwidth and capacity are selected independently by architecture, the models do not necessarily receive identical token inputs.

\paragraph{Simulation architecture selection.}
For the simulation, \texttt{grid\_holdout.py} selects architectures per $(K,N,\mathrm{model})$ on noise-free data. The validation set is a fixed set of 500 independently generated samples held out from training and testing, and each candidate is averaged over three random initializations. Final evaluation uses the selected bandwidth $K{=}32$, repeats training 20 times per noise level, and uses a fixed 2,000-sample test set so all models are compared on identical examples within each noise condition. Each reported 95\% confidence interval is $\bar{x}\pm t_{0.975,19}s/\sqrt{20}$ across the 20 training runs.

\section{Results}
\label{sec:results}

\begin{table}[t]
\centering
\small
\setlength{\tabcolsep}{2.5pt}
\begin{tabular}{lcc}
\toprule
\multicolumn{3}{c}{Shape-based models} \\
\cmidrule(lr){1-3}
$N$ & MLP & Shape Trans. \\
\midrule
50 & 56.3 [50.9, 61.7] & 65.8 [63.4, 68.2] \\
100 & 63.4 [58.2, 68.6] & 66.4 [64.0, 68.7] \\
250 & 69.8 [65.2, 74.5] & 67.5 [65.1, 69.9] \\
1,000 & 69.6 [64.9, 74.4] & 68.5 [63.6, 73.4] \\
\midrule
\multicolumn{3}{c}{Pixel-based models} \\
\cmidrule(lr){1-3}
$N$ & CNN & ViT-Tiny \\
\midrule
50 & 50.1 [49.7, 50.6] & 50.0 [50.0, 50.0] \\
100 & 50.2 [49.8, 50.6] & 50.0 [50.0, 50.0] \\
250 & 51.1 [50.6, 51.5] & 50.0 [50.0, 50.0] \\
1,000 & 52.5 [51.9, 53.1] & 50.0 [50.0, 50.1] \\
\bottomrule
\end{tabular}
\caption{Hybrid simulation on noise-free contours ($\sigma{=}0$), balanced accuracy (\%), mean [95\% CI] over 20 independent training runs. Shape-based columns use $K{=}32$ Fourier harmonics, while pixel-based baselines use the corresponding rasterized shapes.}
\label{tab:sim_clean_ci}
\end{table}

\subsection{BraTS 2020}
\label{sec:results_brats}

\begin{table}[t]
\centering
\small
\setlength{\tabcolsep}{2.8pt}
\begin{tabular}{lrrr}
\toprule
Model & Params & Acc.\ & Bal.\ \\
\midrule
ResNet-18 & 11.2M & 76.8 (11.0) & 65.9 (4.6) \\
ViT-Tiny & 5.43M & 74.6 (20.3) & 63.3 (9.4) \\
\midrule
Shape Trans.\ & 0.80--4.75M & 76.7 (3.2) & 71.0 (3.2) \\
MLP & 2.9--117.3k & \textbf{81.4 (6.0)} & \textbf{71.5 (6.4)} \\
\bottomrule
\end{tabular}
\caption{BraTS 2020 patient-disjoint cross-validation results, mean (SD) in percent. Parameter ranges reflect fold-specific selection.}
\label{tab:brats_acc}
\end{table}

\begin{table}[t]
\centering
\small
\setlength{\tabcolsep}{2.6pt}
\begin{tabular}{lrr}
\toprule
Model & LGG F1 & HGG F1 \\
\midrule
ResNet-18 & 48.4 (11.4) & 84.0 (10.1) \\
ViT-Tiny & 42.2 (24.7) & 78.5 (25.7) \\
\midrule
Shape Trans.\ & 52.8 (9.8) & 84.3 (2.7) \\
MLP & \textbf{54.9 (12.2)} & \textbf{87.7 (5.6)} \\
\bottomrule
\end{tabular}
\caption{BraTS 2020 patient-disjoint per-class F1, mean in percent. Sample standard deviations across outer folds in parentheses.}
\label{tab:brats_f1}
\end{table}

\begin{table}[t]
\centering
\small
\setlength{\tabcolsep}{3pt}
\begin{tabular}{lrrr}
\toprule
Model & Params & Train & Bal.\\
\midrule
ResNet-18 & 11.2M & 1,052\,s & 65.9 \\
ViT-Tiny & 5.43M & 377\,s & 63.3 \\
\midrule
Shape Trans.\ & 0.80--4.75M & 18.1\,s & 71.0 \\
MLP & 2.9--117.3k & 0.16\,s & 71.5 \\
\bottomrule
\end{tabular}
\caption{BraTS 2020 mean final-training wall-clock time and mean balanced accuracy across patient-disjoint folds. Runs use an Apple M3 CPU at batch size 16; selected epoch counts vary by fold.}
\label{tab:brats_efficiency}
\end{table}

\paragraph{Observation.}
The MLP has the highest mean accuracy, balanced accuracy, and LGG F1. Its 71.5\% mean balanced accuracy is 5.6 points above ResNet-18 and 8.2 points above ViT-Tiny. The pooled out-of-fold balanced accuracy is 72.4\%, with a patient-bootstrap 95\% interval of 66.4--77.8\%. The Shape Transformer has higher LGG recall but lower precision and F1; its small balanced-accuracy difference from the MLP is well within fold variability. However, the efficiency difference is much larger as the largest selected MLP is ~8 times smaller than the smallest selected Shape Transformer. Both shape models perform better than the pixel baselines.
\subsection{Controlled Simulation Results}
\label{sec:results_sim}

Tables~\ref{tab:sim_clean_ci} and~\ref{tab:sim_noisy_ci} report balanced accuracy as mean [95\% CI] over 20 independent training runs. The main results use Moindji\'e \etal's direct coefficient-space path, and pixel-based baselines use matched rasters for the same examples. An equivalent spatial-roundtrip implementation gives identical metrics in a fixed-split BraTS check and broadly similar simulation results (Supplementary Sections~C--D).

\paragraph{Observation.}
  On noise-free contours, both shape-based classifiers perform above chance at every sample size, whereas the CNN remains between 50.1\% and 52.5\%. ViT-Tiny
  remains near 50\% with almost no run-to-run variation. A representative rerun at $N{=}250$ confirmed nonzero gradients, but the loss converged near $\log 2$
  and the model predicted a single class for all 2,000 test examples. This behavior indicates optimization collapse so we
  therefore do not interpret the ViT-Tiny result in of itself as evidence of a general disadvantage of pixel-based models.

  The contrast partly reflects the simulation design. Recentering and rescaling remove the explicit deformation variables from the raster even though those
  variables contribute to the label, whereas the deformation token retains them. The MLP attains its highest noise-free mean at $N{=}250$, although its confidence interval is wide. The noise results and attention analysis are reported in the supplement.

  The CNN rises from 50.1--52.5\% without noise to 59.7--66.5\% with noise (Table~\ref{tab:sim_noisy_ci}), a large and counterintuitive change. A possible
  explanation is that vertex-wise contour noise produces raster-level boundary statistics or texture-like cues through its interaction with rasterization,
  consistent with the discussion in Section~\ref{sec:bias}. This mechanism remains a hypothesis and requires direct testing. 
  
  Overall, the deformation-plus-
  shape representation is effective in the noise-free diagnostic. Shape-based models achieve the strongest overall results in this controlled setting, although the unequal information retained by the token and raster
  representations limits broader architecture-level conclusions.

\section{Discussion and Limitations}
\label{sec:discussion}

Our main finding from the BraTS analysis is that an explicit shape-based representation supports competitive or superior class-balanced performance with much smaller models. Under patient-disjoint cross-validation, the MLP has the highest mean balanced accuracy at 71.5\% and the highest mean LGG F1 at 54.9\%. The selected MLPs use 2.9k--117.3k parameters, at least 46 times fewer than the pixel baselines. Its pooled out-of-fold balanced accuracy is 72.4\%, with a patient-bootstrap 95\% interval of 66.4--77.8\%. This is precisely the setting in which we expected a shape-based inductive bias to be valuable, as the segmentation mask is already available and the phenotype is plausibly related to morphology.

The class-wise behavior is asymmetric, which is scientifically useful. Pixel-based baselines remain strong for HGG because the majority class is common and high-grade tumors often have large spatial and size cues. Both shape-based models improve mean LGG F1 over the pixel baselines, suggesting that explicit alignment, scale-aware deformation variables, and harmonic residuals can reduce irrelevant variation before the supervised learner sees the data. This suggests asking where in the class structure, and in which frequency bands, shape analysis provides information that pixel-based models dilute or use inefficiently.

From the simulation, we see that when the functional-shape representation is noise-free and the target is recoverable from a small set of deformation and harmonic coordinates, flattening the token matrix is often enough. In this setting, the coordinate ordering is fixed and directly available to the classifier, so the transformer's input-dependent mixing across tokens provides little additional benefit over a compact MLP. The transformer becomes more useful when the relevant information is distributed across harmonic scales or when interactions between coordinates must adapt to the individual input.

Our representation also addresses two practical goals beyond accuracy. It supports mechanistic interpretability where deformation parameters and harmonic orders can have importance scores and, in future work, ablated. In a biological context, variation within harmonic orders can correspond to variation within organizational orders such as cellular, tissue, and organ level biology. Moreover, it is possible to easily generate synthetic contours based on that information.  It is also scalable for very large imaging collections because each contour becomes a fixed-size matrix of tens or hundreds of numbers, independent of the original image resolution, making storage, CPU training, and inference cheaper than repeated pixel-based model training.

\paragraph{Limitations.}
We reduce each 3D segmentation to one axial slice and one or more connected-component contours, and a multi-slice or surface-based extension would test whether the representation scales to full 3D tumor geometry. BraTS tumour contours are annotated by experts which is a unique feature of this dataset, but using segment-anything-models (SAMs) \cite{Ma2023SegmentAI,carion2026sam3segmentconcepts} in tandem with our method poses an interesting direction for this work which broadens its applicability. Patient grouping prevents subject overlap between outer training and test folds, but 95 LGG contours remain distributed across only five test folds, producing high variability in minority-class estimates. Model selection uses one patient-grouped validation split per outer fold rather than averaging selection over all possible inner folds. We use the first training preshape as the reference $\boldsymbol{\mu}$, leaving training-fold Fr\'echet-mean estimation for future work~\cite{moindjie2025functional}. ResNet-18 and ViT-Tiny are trained from scratch on only a few hundred contours per fold, so their performance may reflect limited-data optimization as well as the input representation, particularly for ViT-Tiny. Binary HGG/LGG grading is simpler than modern molecularly informed glioma classification. Future work can test which harmonic bands remain stable across cohorts and whether attention helps specifically when labels depend on interactions between deformation and local boundary irregularity.

\section{Conclusion}
\label{sec:conclusion}

We define aligned Fourier shape tokens for glioma grading from tumor boundaries. Functional-shape alignment converts each contour into one deformation token plus frequency-ordered residual shape tokens, giving a compact representation of morphology.

The broader implication of our work is that shape analysis can provide substantial dimensionality reduction while retaining mechanistic meaning, with implications beyond medical imaging. Geospatial and remote-sensing data, for example, are often massive, high-dimensional, multiresolution, and spatially structured, and the literature has repeatedly emphasized the computational and statistical challenges of learning directly from these data at scale~\cite{li2016geospatial,zhu2017remote}. Many objects in those settings also have meaningful boundaries, including coastlines, agricultural fields, watersheds, roads, fire perimeters, and urban footprints. For such problems, a shape-based representation could turn very large rasters into compact geometric summaries that can be indexed, audited, and learned from cheaply. Contours may therefore complement full images by providing an initial layer for scalable and interpretable learning whenever morphology is the primary scientific object.

\paragraph{Code.} The implementation uses PyTorch~2.8 for model training and a Python implementation inspired by Moindji\'e \etal's coefficient-space functional-shape alignment. An anonymized repository will accompany double-blind review, and the public URL will be restored after review.

\paragraph{Acknowledgements}

This research was enabled in part by support provided by BC DRI Group and the Digital Research Alliance of Canada (\url{https://alliancecan.ca}).

{
    \small
    \bibliographystyle{ieeenat_fullname}
    \bibliography{main}
}
\clearpage
\newpage

\appendix
\section*{Supplementary Material}
\label{supp:fda}

\subsection*{A. Geometric Meaning of Individual Harmonics}

\begin{figure*}[t]
    \centering
    \includegraphics[width=.75\textwidth]{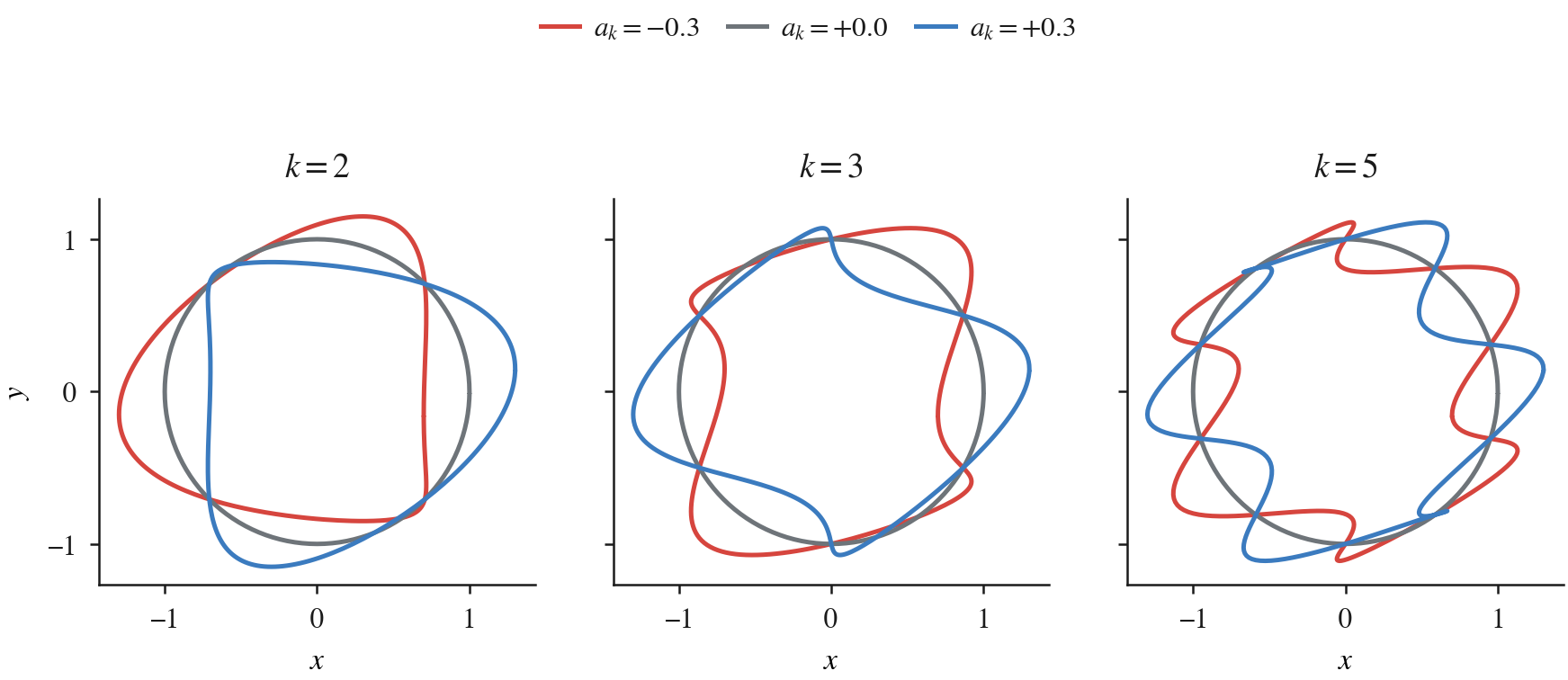}
    \caption{Effect of varying a single Fourier coefficient $a_x$ at different harmonics.}
    \label{fig:warping}
\end{figure*}

Each Fourier harmonic controls a qualitatively different aspect of shape geometry. Figure~\ref{fig:warping} shows the effect of varying a single cosine coefficient $a_x$ at harmonics $k{=}2$, $k{=}3$, and $k{=}5$. Harmonic $k{=}2$ governs bilateral elongation, $k{=}3$ introduces triangularity, and $k{=}5$ adds finer lobes. This frequency-to-geometry correspondence gives our token representation its structure, with each token controlling a specific scale of shape variation.

\subsection*{B. Compute Environment and Timing}

BraTS timing runs were measured on an Apple M3 chip (CPU only). The simulation sweeps and final repeated evaluations were run on Compute Canada cluster resources, so the simulation tables report accuracy and confidence intervals. Table~\ref{tab:timing} reports the fixed-split implementation-check timing, inference time, parameter count, and input dimensionality; patient-grouped final-training times appear in Table~\ref{tab:brats_efficiency}.

\begin{table*}[t]
\centering
\small
\begin{tabular}{llrrcc}
\toprule
Input & Model & Params & Dim. & Train & Infer.\\
\midrule
Pixel & ResNet-18 & 11.2M & $224{\times}224$ & 547\,s & 349.9\,ms \\
Pixel & ViT-Tiny & 5.43M & $224{\times}224$ & 285\,s & 180.2\,ms \\
\midrule
Single & Shape Trans.\ & 3.16M & $9{\times}6$ & 15\,s & 3.66\,ms \\
Single & MLP & 7.5k & $33{\times}6$ & 0.75\,s & 0.03\,ms \\
\midrule
Double & Shape Trans.\ & 3.16M & $9{\times}6$ & 11\,s & 2.97\,ms \\
Double & MLP & 7.5k & $33{\times}6$ & 0.61\,s & 0.02\,ms \\
\bottomrule
\end{tabular}
\caption{Fixed-split implementation-check cost on BraTS 2020 (Apple M3 CPU). Training is wall-clock for 40 epochs at batch size 16. Inference is per batch of 16 samples.}
\label{tab:timing}
\end{table*}

\subsection*{C. Spatial-Roundtrip Robustness Check}

Moindji\'e \etal's formulation remains in functional space after the initial Fourier smoothing. Phase, inverse rotation, and scale are applied directly to the fitted coefficients through the block structure of the basis~\cite{moindjie2025functional}. We call this the \emph{single} path because it requires only one curve-to-Fourier projection. For comparison, we introduced a spatial-roundtrip approximation, which we call the \emph{double} path. It returns to point space, applies the phase shift by cyclic linear interpolation, rotates the sampled curve, and computes a second Fourier projection. The paths are equivalent in the continuous limit, but the double path introduces interpolation, resolution dependence, and a second truncation that can oversmooth the curve.

On the original fixed contour split, the double path gives the same aggregate BraTS results as the direct coefficient-space path for both shape-based classifiers (Table~\ref{tab:brats_roundtrip}). The per-class precision, recall, and F1 values are also identical across paths. The simulation results are likewise comparable, with overlapping confidence intervals at every sample size (Table~\ref{tab:sim_clean_roundtrip} and Supplementary Section~D). We therefore retain the direct coefficient-space path in the patient-grouped main results and treat the double path as an implementation robustness check.

\begin{table*}[t]
\centering
\small
\setlength{\tabcolsep}{3pt}
\begin{tabular}{llrrrrr}
\toprule
Path & Model & Acc. & Bal. & G-mean & LGG & HGG \\
\midrule
Single & Shape Trans. & 85.3 & 72.0 & 67.8 & 47.8 & 96.2 \\
Single & MLP & 85.3 & 76.6 & 75.0 & 60.9 & 92.4 \\
Double & Shape Trans. & 85.3 & 72.0 & 67.8 & 47.8 & 96.2 \\
Double & MLP & 85.3 & 76.6 & 75.0 & 60.9 & 92.4 \\
\bottomrule
\end{tabular}
\caption{Fixed-split BraTS 2020 comparison of the direct coefficient-space (single) and spatial-roundtrip (double) implementations. LGG/HGG columns report class recall, and G-mean is their geometric mean.}
\label{tab:brats_roundtrip}
\end{table*}

\begin{table*}[t]
\centering
\small
\setlength{\tabcolsep}{3pt}
\begin{tabular}{lcccc}
\toprule
$N$ & MLP-S & Shape-S & MLP-D & Shape-D \\
\midrule
50 & 56.3 [50.9, 61.7] & 65.8 [63.4, 68.2] & 58.4 [52.5, 64.3] & 66.3 [65.0, 67.7] \\
100 & 63.4 [58.2, 68.6] & 66.4 [64.0, 68.7] & 66.3 [61.9, 70.8] & 67.8 [66.1, 69.6] \\
250 & 69.8 [65.2, 74.5] & 67.5 [65.1, 69.9] & 71.5 [67.2, 75.8] & 67.8 [65.5, 70.1] \\
1,000 & 69.6 [64.9, 74.4] & 68.5 [63.6, 73.4] & 68.4 [64.2, 72.6] & 66.8 [62.8, 70.8] \\
\bottomrule
\end{tabular}
\caption{Noise-free hybrid simulation comparison of the direct coefficient-space (S) and spatial-roundtrip (D) implementations, balanced accuracy (\%), mean [95\% CI] over 20 independent training runs.}
\label{tab:sim_clean_roundtrip}
\end{table*}

\subsection*{D. Simulation with Noise}
\begin{table*}[t]
\centering
\small
\setlength{\tabcolsep}{2.5pt}
\begin{tabular}{lcccccc}
\toprule
$N$ & MLP-S & Shape-S & MLP-D & Shape-D & CNN & ViT-Tiny \\
\midrule
50 & 58.7 [51.2, 66.3] & 66.9 [63.5, 70.3] & 61.0 [53.4, 68.5] & 69.4 [67.9, 70.9] & 59.7 [57.1, 62.2] & 50.0 [50.0, 50.0] \\
100 & 57.8 [50.5, 65.2] & 66.5 [64.2, 68.8] & 56.1 [49.1, 63.1] & 68.8 [66.7, 71.0] & 66.4 [64.5, 68.2] & 50.0 [49.9, 50.0] \\
250 & 54.8 [47.5, 62.2] & 67.6 [65.0, 70.2] & 59.0 [52.3, 65.7] & 68.0 [66.1, 70.0] & 66.5 [64.8, 68.3] & 50.5 [49.5, 51.5] \\
1,000 & 59.9 [54.4, 65.5] & 65.6 [62.3, 68.9] & 60.8 [54.6, 67.1] & 68.9 [66.4, 71.4] & 64.6 [62.8, 66.5] & 52.8 [50.3, 55.3] \\
\bottomrule
\end{tabular}
\caption{Hybrid simulation with contour vertex noise ($\sigma{=}0.10$), balanced accuracy (\%), mean [95\% CI] over 20 independent training runs. Shape-based and pixel-based classifiers are compared on the same generated examples.}
\label{tab:sim_noisy_ci}
\end{table*}
\subsection*{E. Simulation Attention by Noise and Tokenization Path}

\begin{figure*}[t]
    \centering
    \includegraphics[width=.99\textwidth]{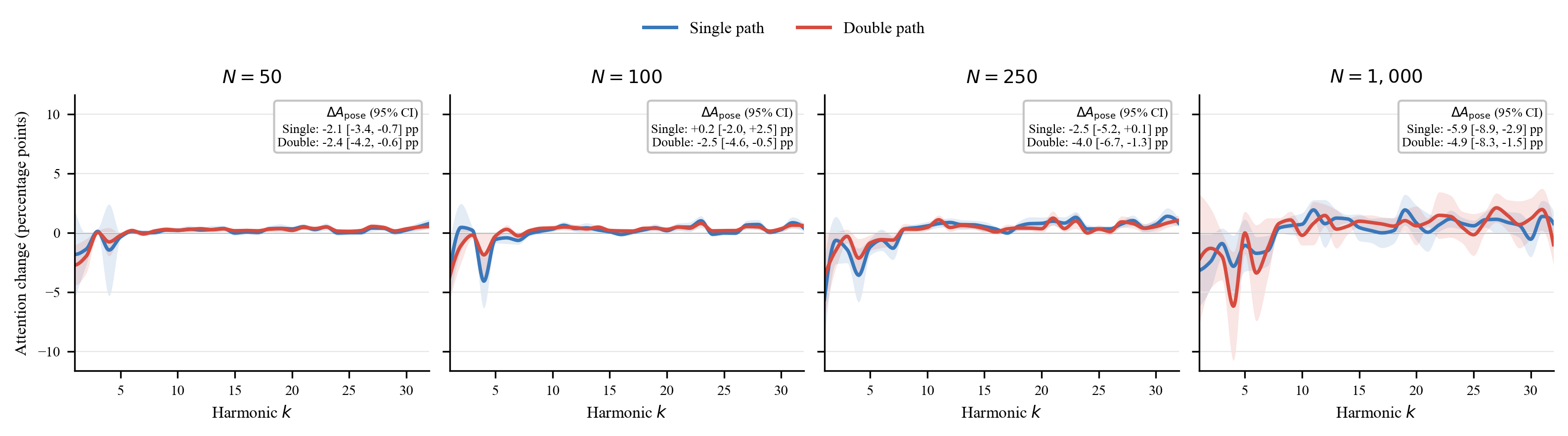}
    \caption{Paired noisy-minus-clean change in pose-query attention to harmonics, in attention percentage points, for the single and double tokenization paths across sample sizes. Both paths are overlaid in each sample-size panel. Solid curves are means over 20 matched seeds and shaded regions are 95\% Student-$t$ confidence intervals. Curves use shape-preserving interpolation for display; summaries are computed at the original integer harmonics. Each annotation gives the paired percentage-point change in pose-token self-attention and its 95\% confidence interval for both paths. Models were trained separately at each noise level.}
    \label{fig:harmonic_attention}
\end{figure*}

Figure~\ref{fig:harmonic_attention} shows the paired change in first-layer Shape Transformer attention from the noise-free condition to $\sigma{=}0.10$ for the single and double tokenization paths at each reported sample size. Each curve is noisy-minus-clean attention from the pose-token query to each harmonic-token key, expressed in attention percentage points, paired by training seed, averaged first over heads and the fixed 2,000-example test set, and then summarized over the same 20 seeds used in Tables~\ref{tab:sim_clean_ci} and~\ref{tab:sim_noisy_ci}. Positive values indicate that noise shifts more pose-query attention toward a harmonic; negative values indicate a reduction. Pose-token self-attention decreases significantly under noise for the double path at every sample size. For the single path, the decrease is significant at $N{=}50$ and $N{=}1{,}000$; the intervals at $N{=}100$ and $N{=}250$ include zero. These shifts describe changes in learned attention allocation and should not by themselves be interpreted as causal feature importance.

\subsection*{F. Simulated Shape Examples}

\begin{figure*}[t]
    \centering
    \includegraphics[width=.78\textwidth]{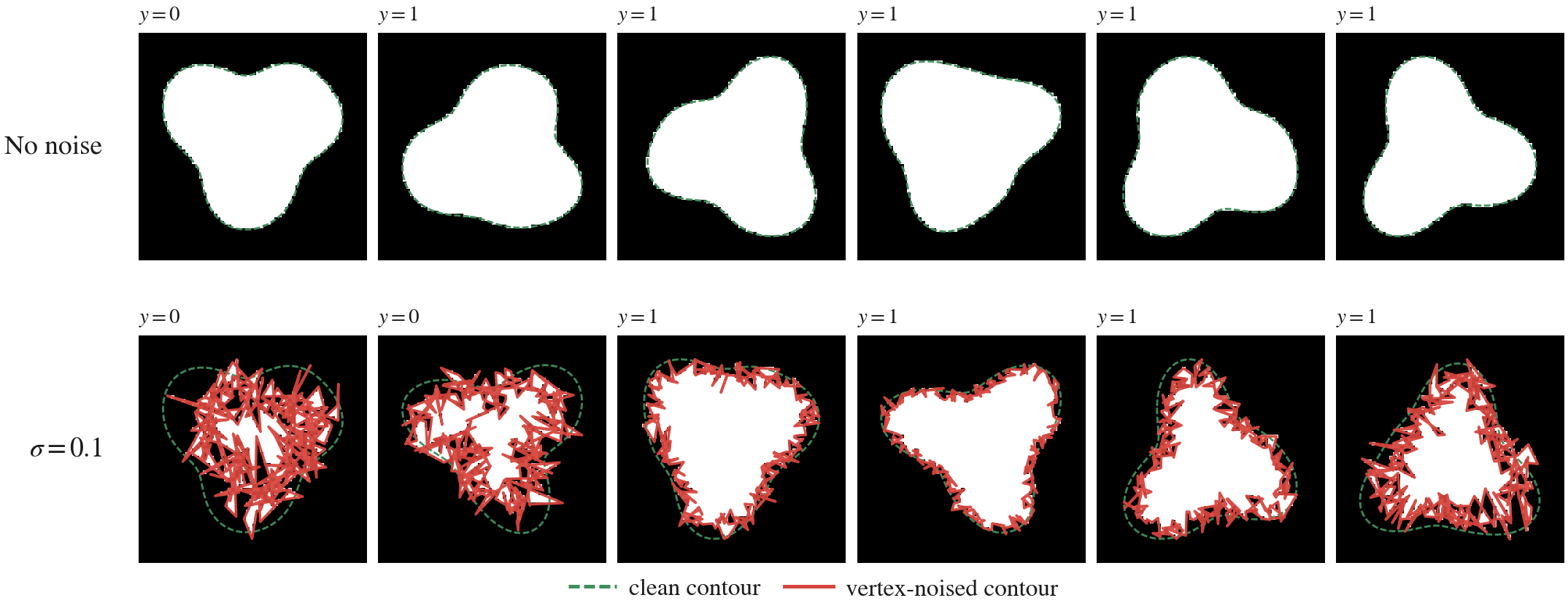}
    \caption{Example simulated shapes. Each cell shows a $64{\times}64$ raster with the noise-free contour (green dashed) and the vertex-noised contour (red). Labels combine deformation evidence from scale and translation with class-dependent mid-frequency bumps; pixel-based models read the raster, while shape-based models read the deformation and residual shape tokens.}
    \label{fig:sim_rasters}
\end{figure*}

\subsection*{G. Full Hyperparameter-Sweep Tables}

The MLP and Shape Transformer sweeps on BraTS cover the reported token-classifier configurations. The full per-configuration logs are in the code repository. The completed hybrid simulation aggregates are stored under \texttt{results/hybrid\_sim/}, including \texttt{sim\_single\_ci\_results.json}, \texttt{sim\_double\_ci\_results.json}, and \texttt{pixel\_ci\_results.json}.

\end{document}